\documentclass[aps,pre,twocolumn,groupedaddress,amsmath,showpacs,eqsecnum,amssymb]{revtex4}
    \usepackage{dcolumn}
    \usepackage{bm}
    \usepackage{graphicx}
    \usepackage{dcolumn}
\newcommand\beq{\begin{equation}}
\newcommand\eeq{\end{equation}}
\newcommand\beqa{\begin{eqnarray}}
\newcommand\eeqa{\end{eqnarray}}
\newcommand{\nn}{\nonumber\\}

        \newcommand\kk[1]{{{{{#1}}}}}
\begin{document}


\title{First-order Chapman--Enskog velocity distribution function in  a granular gas}


\author{Jos\'e Mar\'{\i}a Montanero}
\email{jmm@unex.es} \affiliation{Departamento de Electr\'onica e
Ingenier\'{\i}a Electromec\'anica, Universidad de Extremadura,
E-06071 Badajoz, Spain}
\author{Andr\'es Santos}
\email{andres@unex.es}
\homepage{http://www.unex.es/eweb/fisteor/andres/}
\author{Vicente Garz\'o}
\email{vicenteg@unex.es}
\homepage{http://www.unex.es/eweb/fisteor/vicente/}
\affiliation{Departamento de F\'{\i}sica, Universidad de
Extremadura, E-06071 Badajoz, Spain}


\date{\today}

\begin{abstract}
A method is devised to measure the first-order Chapman--Enskog
velocity distribution function associated with the heat flux in a
dilute granular gas. The method is based on the application of a
homogeneous, anisotropic velocity-dependent external force which
produces  heat flux in the absence of gradients. The form of the
force is found under the condition that, in the linear response
regime, the deviation of the velocity distribution function from
that of the homogeneous cooling state obeys the same linear integral
equation as the one derived from the conventional Chapman--Enskog
expansion. The Direct Simulation Monte Carlo method is used to solve
the corresponding Boltzmann equation and measure the dependence of
the (modified) thermal conductivity  on the coefficient of normal
restitution $\alpha$. Comparison with previous simulation data
obtained from the Green--Kubo relations [Brey et al., J. Phys.:
Condens. Matter \textbf{17}, S2489 (2005)] shows an excellent
agreement, both methods consistently showing that  the first Sonine
approximation dramatically overestimates  the thermal conductivity
for high inelasticity ($\alpha\lesssim 0.7$). Since our method is
tied to the Boltzmann equation, the results indicate that the
failure of the first Sonine approximation is not due to velocity
correlation effects absent in the Boltzmann framework. This is
further confirmed by an analysis of the first-order Chapman--Enskog
velocity distribution function and its three first Sonine
coefficients obtained from the simulations.

\end{abstract}



\maketitle

\section{Introduction}
It is well established that granular fluids can be successfully
described by fluid dynamics, nonequilibrium statistical mechanics,
and kinetic theory conveniently modified to account for the energy
dissipation due to the inelasticity of collisions \cite{C90}. In
particular, the Navier--Stokes (NS) transport coefficients
characterize the departure from homogeneous situations due to small
spatial gradients of the hydrodynamic fields (density, flow
velocity, and granular temperature). In the case of a monodisperse
granular fluid, the relevant transport coefficients are (i) the
shear and the bulk viscosities (which relate the pressure tensor to
the flow velocity gradients) and (ii) the thermal conductivity
$\kappa$ and a  coefficient $\mu$ not present for elastic collisions
(which relate the heat flux to the temperature and density
gradients, respectively).

By using statistical-mechanical methods, formal expressions for the
transport coefficients in the form of Green--Kubo (GK) relations
have been recently derived
\cite{GvN00,DG01,DBL02,DB02,BDRM03,DBB05}. While these GK relations
have a structure formally similar to their elastic counterparts,
they are not simple extensions of the latter and expose the
subtleties associated with the energy dissipation in collisions. In
principle, the GK formulae can be used to get the transport
coefficients by measuring the appropriate time correlation functions
in the homogeneous cooling state (HCS) from computer simulations. In
the low-density regime, and assuming the validity of the molecular
chaos hypothesis, a kinetic theory description based on the
Boltzmann equation is suitable \cite{GS95,BDS97}. In that case, the
GK relations adopt a more explicit form, where the generator of
dynamics is the linearized Boltzmann operator. These low-density GK
expressions have been recently employed in computer simulations by
means of the Direct Simulation Monte Carlo (DSMC) method to evaluate
the dependence of the transport coefficients on inelasticity
\cite{BRM04,BRMMG05}.

In a kinetic theory description, an alternative route to determine
the transport coefficients is provided by the Chapman--Enskog (CE)
method \cite{CC70}. Thus, the NS coefficients have been derived from
the Boltzmann equation for monodisperse \cite{BDKS98,BC01,GM02} and
multicomponent  \cite{GD02} dilute gases, as well as from the Enskog
equation for moderately dense gases \cite{GD99,L05}. The equivalence
between the GK and CE expressions derived from the Boltzmann
equation has been checked for several transport coefficients
\cite{DB02,DG01,DBL02}. In the CE method, the transport coefficients
are given in terms of the solutions of linear integral equations
involving the linearized collision operator. Since the solutions are
not exactly known, even in the elastic case, one usually resorts to
the so-called Sonine approximations. This allows one to get explicit
expressions for the transport coefficients in terms of the
parameters of the system.

As in the elastic case, the first Sonine approximation to the shear
viscosity $\eta$ is seen to agree quite well, even for strong
dissipation, with DSMC results obtained from three independent
methods: (i)  the time decay of a weak transverse shear wave in the
HCS \cite{BRMC99}, (ii) the GK method \cite{BRM04,BRMMG05}, and
(iii) a modified simple shear flow \cite{GM02,GM03,MSG05}.
Concerning the heat transport coefficients $\kappa$ and $\mu$, a
good agreement between the first Sonine predictions and DSMC
simulations of the corresponding GK formulae has been observed for
values of the  coefficient of normal restitution $\alpha\gtrsim 0.7$
\cite{BRM04,BRMMG05}. However, significant discrepancies appear for
stronger dissipation ($\alpha\lesssim 0.7$). While the range
$\alpha\gtrsim 0.7$ widely covers those situations of practical or
experimental interest,  it is still important, from a fundamental
point of view, to understand the origin of those discrepancies for
$\alpha\lesssim 0.7$. \kk{This can shed light on the physical
mechanisms playing a relevant role in the dynamics of granular flows
at strong dissipation.}

Three  scenarios are possible to explain the above discrepancies:
(i) although the first Sonine approximation might accurately
estimate the true coefficients $\kappa$ and $\mu$ described by the
Boltzmann equation, the HCS exhibits relevant velocity correlations
for strong inelasticities, even in the low-density limit, not
accounted for by the inelastic Boltzmann equation; (ii) the
discrepancies are an artifact of the DSMC method when applied to the
evaluation of the time correlation functions needed in the GK
representation; (iii) the disagreement is simply due to the
limitations for high dissipation of the first Sonine approximation
because of the deviation of the HCS velocity distribution from its
Gaussian form.

The aim of this paper is to contribute to the clarification of the
above controversial issue. Specifically, we want to confirm or
discard the possibility (iii) mentioned above. To that end, we will
work within the framework of the Boltzmann equation and will use the
DSMC method to compute \textit{one-time} averages, in contrast to
the GK route which involves \textit{two-time} averages. Our method
consists of perturbing the HCS by the application of a \textit{weak}
non-conservative (i.e., velocity-dependent) external force which
produces a heat flux in the absence of inhomogeneities. This force,
which mimics the effect of a thermal gradient, must be chosen under
the condition that the perturbed velocity distribution function to
first order coincides exactly with the one derived from the CE
method at the NS order. This method is a \kk{non-trivial} extension
to the granular case of the one devised time ago by Evans \cite{E82}
and, independently, by Gillan and Dixon \cite{GD83}. \kk{However,
while in the elastic case the force is parallel to the heat flux,
this is not so in the inelastic case, as a consequence of the
non-Gaussian form of the HCS.} For elastic systems, the method has
been successfully applied to get the thermal conductivity of
Lennard--Jones \cite{E86} and hard-sphere \cite{MS97} fluids; the
corresponding nonlinear problem has also been studied
\cite{L89,GS91}. Apart from measuring the transport coefficients,
the method allows one to get the velocity distribution function to
first order in the external field and this is one of the main
objectives of this paper.

As will be seen later on, in order to determine the thermal
conductivity $\kappa$ and its associated velocity distribution, the
application of an external force is not enough since an additional
stochastic term is needed, \kk{which} complicates the implementation
of the simulation method. The situation is even worse in the case of
$\mu$ since it is coupled to $\kappa$. Nevertheless, the integral
equation associated with the coefficient $\kappa'\equiv
\kappa-n\mu/2T$ can actually be simulated by the action of an
external force only. Therefore, in this paper we focus on this
modified thermal conductivity coefficient $\kappa'$ and its
corresponding velocity distribution. The simulation results obtained
from the present method for $\kappa'$ agree well with those obtained
from the alternative GK method \cite{BRMMG05}. Moreover, the
velocity distribution function for high inelasticity differs
appreciably from the Sonine expansion truncated after the first
term. These results strongly support the scenario (iii) mentioned
before. In fact, a modified first Sonine approximation, where the
Gaussian weight appearing in the Sonine expansion is replaced by the
HCS distribution, compares fairly well with the simulation data for
the whole range of inelasticities \cite{GS06}.

The organization of the paper is as follows. The Boltzmann equation
for inelastic hard spheres and its solution provided by the CE
method is presented in Sec.\ \ref{sec2}. The linear integral
equations for the NS distributions associated with the heat flux are
worked out and the modified thermal conductivity coefficient
$\kappa'$ is introduced. In Sec.\ \ref{sec3} we propose the method
to get the first-order (NS) distributions by the application of
velocity-dependent external forces in spatially uniform states. The
numerical results obtained from the DSMC method for $\kappa'$ and
its corresponding velocity distribution function are presented and
discussed in Sec.\ \ref{sec4}. The paper is closed with the
conclusions in Sec.\ \ref{sec5}.

\section{Inelastic Boltzmann equation and Chapman--Enskog expansion\label{sec2}}
For the sake of completeness and to fix the notation, in this
Section we summarize the main known results obtained by applying the
CE method to the Boltzmann equation.

Let us consider a granular gas composed by smooth inelastic disks
($d=2$) or spheres ($d=3$) of mass $m$ and diameter $\sigma$. The
inelasticity of collisions among all pairs is characterized by a
constant coefficient of normal restitution $\alpha\leq 1$. In the
low-density regime, the evolution of the one-particle velocity
distribution function $f({\bf r},{\bf v},t)$ is given by the
Boltzmann kinetic equation \cite{GS95,BDS97}. In the absence of
external forces, this equation reads, \kk{in standard notation,}
\begin{equation}
\label{2.1}
\left(\partial _{t}+{\bf v}\cdot \nabla \right)f({\bf r},{\bf v};t)
=J\left[{\bf v}|f,f\right] ,
\end{equation}
where  $J[{\bf v}|f,f]$ is the Boltzmann collision operator
\cite{GS95,BDS97}. Collisions conserve mass and momentum, but energy
is dissipated (except, of course, if $\alpha=1$):
\beq
\int d\mathbf{v}\left\{
\begin{array}{c}
1\\
\mathbf{v}\\
V^2
\end{array}\right\}
J[\mathbf{v}|f,f]=\left\{
\begin{array}{c}
0\\
\mathbf{0}\\
-\frac{d}{m}\zeta nT
\end{array}\right\}.
\label{n1}
\eeq
Here
\beq
n=\int d\mathbf{v} f(\mathbf{v})
\label{n2}
\eeq
is the number density, $\mathbf{V}=\mathbf{v}-\mathbf{u}$ is the
peculiar velocity, where
\beq
\mathbf{u}=\frac{1}{n}\int d\mathbf{v} f(\mathbf{v})
\label{n3}
\eeq
is the flow velocity,
\beq
T=\frac{m}{dn}\int d\mathbf{v} V^2f(\mathbf{v})
\label{n4}
\eeq
is the granular temperature, and $\zeta$ is the cooling rate.
Similarly, the fluxes can be obtained as moments of the velocity
distribution function. In particular, the irreversible part of the
pressure tensor is
\beq
\Pi_{ij}=m\int d{\bf v}
\left(V_iV_j-\frac{1}{d}V^2\delta_{ij}\right)f({\bf v})
\label{2.8bis}
\end{equation}
and the heat flux is
\begin{equation}
{\bf q}=\frac{m}{2}\int d{\bf v} V^{2}{\bf V} f({\bf v}).
\label{2.8}
\end{equation}

The CE method assumes the existence of a {\em normal} solution in
which all the space and time dependence of the distribution function
appears through a functional dependence on the hydrodynamic fields.
For small spatial variations, this functional dependence can be made
local in space and time through an expansion in gradients of the
fields: $f=f^{(0)}+f^{(1)}+\cdots$.
 The local reference state $f^{(0)}$ is
constrained to have the same first few moments
(\ref{n2})--(\ref{n4}) as the exact distribution $f$. The kinetic
equation for $f^{(0)}$ is \cite{BDKS98}
\beq
\frac{1}{2}\zeta^{(0)}\frac{\partial}{\partial \mathbf{V}}\cdot
\left(\mathbf{V}f^{(0)}\right)=J[\mathbf{V}|f^{(0)},f^{(0)}],
\label{n6}
\eeq
where $\zeta^{(0)}$ is defined by setting $f\to f^{(0)}$ in Eq.\
(\ref{n1}). Equation (\ref{n6}) can be recognized as the one
satisfied by the HCS \cite{GS95,vNE98}, parameterized by the
\textit{local} values of $n$, $\mathbf{u}$, and $T$. The exact
solution to Eq.\ (\ref{n6}) has not been found, although some of its
properties are known \cite{vNE98,EP97}.

Since $f^{(0)}(\mathbf{V})$ is an isotropic function, its Sonine
expansion is
\beq
f^{(0)}(\mathbf{V})= f_M(\mathbf{V})\left[1+\sum_{k=2}^\infty a_k
L_k^{(\frac{d-2}{2})}(c^2) \right],
\label{n2.2}
\eeq
where
\beq
f_M(\mathbf{V})=n v_0^{-d}\pi^{-d/2}e^{-c^2}
\label{n2.3}
\eeq
is the Maxwellian distribution,
\beq
\mathbf{c}\equiv \frac{\mathbf{V}}{v_0(T)}
\label{n2.2bis}
\eeq
is the velocity relative to the thermal speed $v_0(T)\equiv
\sqrt{2T/m}$ and $L_k^{(p)}(x)$ are the generalized Laguerre
polynomials whose orthogonality relation is \cite{AS72}
\beq
\int_0^\infty dx\, x^{p} e^{-x}L_k^{(p)}(x)
L_\ell^{(p)}(x)=\frac{\Gamma(k+1+p)}{k!}\delta_{k\ell}.
\label{5.4}
\eeq
The coefficients $a_k$ are related to the velocity moments,
measuring the deviation of $f^{(0)}$ from the Gaussian. In
particular, $a_2$ is the fourth cumulant (or kurtosis) of the
distribution function $f^{(0)}$:
\beq
\int d\mathbf{V}V^4
f^{(0)}(\mathbf{V})=d(d+2)\frac{nT^2}{m^2}(1+a_2).
\label{n2.4}
\eeq
By inserting Eq.\ (\ref{n2.2}) into Eq.\ (\ref{n1}) and neglecting
 $a_k$ with $k\geq 3$ and nonlinear terms in $a_2$, one gets
 \beq
\zeta^{(0)}=\frac{d+2}{4d}\nu_0(1-\alpha^2)\left(1+\frac{3}{16}a_2\right),
\label{x2.29}
\eeq
where
\begin{equation}
\nu_0= \frac{8}{d+2}\frac{\pi^{(d-1)/2}}{\Gamma(d/2)}n\sigma
^{d-1}\left( \frac{T}{m} \right) ^{1/2}
\label{x3.14b}
\end{equation}
is an effective collision frequency.
 \kk{An excellent} estimate of $a_2$ is \cite{MS00,CDPT03}
\beq
a_2= \frac{16(1-\alpha)(1-2\alpha^2)}{25+24d-\alpha (57-
8d)-2(1-\alpha)\alpha^2}.
\label{n2.5}
\eeq

To first order, the application of the CE method yields the linear
integral equation \cite{BDKS98}
\beq
\left(\partial_t^{(0)}
+\mathcal{L}\right)f^{(1)}(\mathbf{V})=\mathbf{A}(\mathbf{V})\cdot\nabla
\ln T+\mathbf{B}(\mathbf{V})\cdot\nabla \ln n,
\label{1}
\eeq
where we have particularized to the case $\nabla_i {u}_j=0$. In Eq.\
(\ref{1}), $\partial_t^{(0)}$ is an operator acting on any function
of temperature as
\beq
\partial_t^{(0)} X(T)=\left(\partial_t^{(0)} T\right) \frac{\partial X(T)}{\partial T}=-\zeta^{(0)}T\frac{\partial X(T)}{\partial T},
\label{n2.7}
\eeq
$\mathcal{L}$ is the linearized Boltzmann collision operator defined
as
\beq
\mathcal{L}X(\mathbf{V})=-J[\mathbf{V}|f^{(0)},X]-J[\mathbf{V}|X,f^{(0)}],
\label{n2.6}
\eeq
and
\beq
\mathbf{A}(\mathbf{V})\equiv
\frac{1}{2}\mathbf{V}\frac{\partial}{\partial\mathbf{V}}\cdot
\left[\mathbf{V}f^{(0)}(\mathbf{V})\right]-\frac{T}{m}\frac{\partial}{\partial\mathbf{V}}f^{(0)}(\mathbf{V}),
\label{2}
\eeq
\beq
\mathbf{B}(\mathbf{V})\equiv
-\mathbf{V}f^{(0)}(\mathbf{V})-\frac{T}{m}\frac{\partial}{\partial\mathbf{V}}f^{(0)}(\mathbf{V}).
\label{3}
\eeq
The structure of the solution to Eq.\ (\ref{1}) is
\beq
f^{(1)}(\mathbf{V})=\boldsymbol{\mathcal{A}}(\mathbf{V})\cdot\nabla\ln
T+ \boldsymbol{\mathcal{B}}(\mathbf{V})\cdot\nabla\ln n.
\label{n2.8}
\eeq
Taking into account Eq.\ (\ref{2.8}), the heat flux to first order
(NS order) is
\beq
\mathbf{q}^{(1)}=-\kappa \nabla T-\mu\nabla n,
\label{n2.14}
\eeq
where
\beq
\kappa=-\frac{1}{dT}\int
d\mathbf{V}\mathbf{S}(\mathbf{V})\cdot\boldsymbol{\mathcal{A}}(\mathbf{V})
\label{n2.15}
\eeq
is the thermal conductivity coefficient and
\beq
\mu=-\frac{1}{dn}\int
d\mathbf{V}\mathbf{S}(\mathbf{V})\cdot\boldsymbol{\mathcal{B}}(\mathbf{V})
\label{n2.16}
\eeq
is a  new coefficient absent for normal gases. In Eqs.\
(\ref{n2.15}) and (\ref{n2.16}) we have introduced the function
\beq
\mathbf{S}(\mathbf{V})=\left(\frac{m}{2}V^2-\frac{d+2}{2}T\right)\mathbf{V}.
\label{n2.16bis}
\eeq

By dimensional analysis,
$\boldsymbol{\mathcal{A}}(\mathbf{V})=nv_0^{-d}\lambda
\boldsymbol{\mathcal{A}}^*(\mathbf{c})$, where $\lambda\sim 1/n
\sigma^{d-1}$ is the mean free path and
$\boldsymbol{\mathcal{A}}^*(\mathbf{c})$ is a dimensionless function
of the reduced velocity $\mathbf{c}$ defined in Eq.\
(\ref{n2.2bis}). A similar relation holds for
$\boldsymbol{\mathcal{B}}(\mathbf{V})$. Consequently,
\beqa
\partial_t^{(0)} f^{(1)}(\mathbf{V})&=&\frac{1}{2}\zeta^{(0)}\frac{\partial}{\partial\mathbf{V}}\cdot
\left[\mathbf{V}f^{(1)}(\mathbf{V})\right]\nn &&-
\zeta^{(0)}\boldsymbol{\mathcal{A}}(\mathbf{V})\cdot \left(\nabla\ln
n+\frac{1}{2}\nabla\ln T\right).\nn&&
\label{9}
\eeqa
Equating the coefficients of $\nabla\ln T$ and $\nabla\ln n$ in Eq.\
(\ref{1}), one obtains the following pair of linear integral
equations:
\beq
\left(\mathcal{L}+\frac{1}{2}\zeta^{(0)}\frac{\partial}{\partial\mathbf{v}}\cdot
\mathbf{V}-\frac{1}{2}\zeta^{(0)}\right)\boldsymbol{\mathcal{A}}(\mathbf{V})=\boldsymbol{{A}}(\mathbf{V}),
\label{n2.11}
\eeq
\beq
\left(\mathcal{L}+\frac{1}{2}\zeta^{(0)}\frac{\partial}{\partial\mathbf{v}}\cdot
\mathbf{V}\right)\boldsymbol{\mathcal{B}}(\mathbf{V})-\zeta^{(0)}\boldsymbol{\mathcal{A}}(\mathbf{V})=\boldsymbol{{B}}(\mathbf{V}).
\label{n2.12}
\eeq
While Eq.\ (\ref{n2.11}) is a closed equation for the unknown
function $\boldsymbol{\mathcal{A}}$, Eq.\ (\ref{n2.12}) is coupled
to Eq.\ (\ref{n2.11}), so that one needs first to know
$\boldsymbol{\mathcal{A}}$ to determine $\boldsymbol{\mathcal{B}}$.
However, the combination $\boldsymbol{\mathcal{A}}'\equiv
\boldsymbol{\mathcal{A}}-\frac{1}{2}\boldsymbol{\mathcal{B}}$
verifies the closed integral equation
\beq
\left(\mathcal{L}+\frac{1}{2}\zeta^{(0)}\frac{\partial}{\partial\mathbf{v}}\cdot
\mathbf{V}\right)\boldsymbol{\mathcal{A}}'(\mathbf{V})=\boldsymbol{{A}}'(\mathbf{V}),
\label{n2.13}
\eeq
where
\beqa
\boldsymbol{{A}}'(\mathbf{V})&\equiv&
\boldsymbol{{A}}(\mathbf{V})-\frac{1}{2}\boldsymbol{{B}}(\mathbf{V})\nn
&=& -\frac{\partial}{\partial\mathbf{V}}\cdot
\left[\mathsf{G}'(\mathbf{V})f^{(0)}(\mathbf{V})\right],
\label{n2.13bis}
\eeqa
and we have called
\beq
G_{ij}'(\mathbf{V})\equiv \frac{T}{2m}\delta_{ij}-\frac{1}{2}V_iV_j.
\label{n2.22}
\eeq
\kk{Thus, the function $\boldsymbol{{A}}'(\mathbf{V})$ can be
expressed as the divergence of a tensor. Let us see that the same
applies to the function $\boldsymbol{{A}}(\mathbf{V})$. Note first
the identity
\beq
\mathbf{V}\frac{\partial}{\partial\mathbf{V}}\cdot
\left[\mathbf{V}f^{(0)}(\mathbf{V})\right]=-\mathbf{V}f^{(0)}(\mathbf{V})+
\frac{\partial}{\partial\mathbf{V}}\cdot
\left[\mathbf{V}\mathbf{V}f^{(0)}(\mathbf{V})\right].
\label{5}
\eeq
Further, taking into account that   $f^{(0)}(\mathbf{V})$ is an
\textit{isotropic} function, we can write
\beq
\mathbf{V}\frac{\partial}{\partial\mathbf{V}}\cdot
\left[\mathbf{V}f^{(0)}(\mathbf{V})\right]=(d-2)\mathbf{V}f^{(0)}(\mathbf{V})+
\frac{\partial}{\partial\mathbf{V}}\left[{V^2}f^{(0)}(\mathbf{V})\right].
\label{4}
\eeq
Equating the right-hand sides of Eqs.\ (\ref{5}) and (\ref{4}), one
has
\beq
V_i f^{(0)}(\mathbf{V})=\frac{1}{d-1}\frac{\partial}{\partial
V_j}\left[\left(V_iV_j-V^2\delta_{ij}\right)f^{(0)}(\mathbf{V})\right].
\label{18}
\eeq
Insertion of this into Eq.\ (\ref{2}) yields
\beq
\mathbf{A}(\mathbf{V})=-\frac{\partial}{\partial\mathbf{V}}\cdot
\left[\mathsf{G}(\mathbf{V})f^{(0)}(\mathbf{V})\right],
\label{n2.21a}
\eeq
where the tensor $\mathsf{G}(\mathbf{V})$ is
\beq
G_{ij}(\mathbf{V})=
-\left[\frac{V^2}{2(d-1)}-\frac{T}{m}\right]\delta_{ij}-\frac{d-2}{2(d-1)}V_i
V_j.
\label{8}
\eeq
}

 The problem of determining the functions
$\boldsymbol{\mathcal{A}}(\mathbf{V})$ and
$\boldsymbol{\mathcal{B}}(\mathbf{V})$ from Eqs.\ (\ref{n2.11}) and
(\ref{n2.12}) is fully equivalent to that of determining the
functions $\boldsymbol{\mathcal{A}}(\mathbf{V})$ and
$\boldsymbol{\mathcal{A}}'(\mathbf{V})$ from Eqs.\ (\ref{n2.11}) and
(\ref{n2.13}), respectively. The latter will be the point of view
adopted in the remainder of this Section as well as in Sec.\
\ref{sec3}. In particular, the coefficient $\mu$ defined by Eq.\
(\ref{n2.16}) can be expressed as
\beq
\mu=\frac{2T}{n}(\kappa-\kappa'),
\label{n2.17}
\eeq
where
\beq
\kappa'=-\frac{1}{dT}\int
d\mathbf{V}\mathbf{S}(\mathbf{V})\cdot\boldsymbol{\mathcal{A}}'(\mathbf{V})
\label{n2.18}
\eeq
is a \textit{modified} thermal conductivity coefficient. In terms of
it, Eq.\ (\ref{n2.14}) can be rewritten as
\beq
\mathbf{q}^{(1)}=-\kappa' \nabla T-\mu T^{-1/2}\nabla
\left(nT^{1/2}\right),
\label{n2.14bis}
\eeq
Therefore, one has  $\mathbf{q}^{(1)}=-\kappa \nabla T$ in those
points where the density gradient vanishes, while
$\mathbf{q}^{(1)}=-\kappa' \nabla T$ in those points where the
gradient of the \kk{cooling rate} (which is proportional to
$nT^{1/2}$) vanishes. In this context, it is worth mentioning that
one of the hydrodynamic modes of a dilute granular gas corresponds
to an excitation where $\zeta^{(0)}=\text{const}$ at zero flow
velocity \cite{BD05}. In the elastic case, $f^{(0)}$ is the
Gaussian, $\zeta^{(0)}=0$, and
$\mathbf{A}(\mathbf{V})=\mathbf{A}'(\mathbf{V})$, so that
$\kappa=\kappa'$ and $\mu=0$.

 Formal expressions for $\kappa$ and $\kappa'$
can be derived   by multiplying both sides of Eqs.\ (\ref{n2.11})
and (\ref{n2.13}) by $\mathbf{S}(\mathbf{V})$ and integrating over
velocity. The results are
\beq
\kappa=\frac{d+2}{2}\frac{nT}{m}\frac{1+2a_2}{\nu_\kappa-2\zeta^{(0)}},
\label{n2.19}
\eeq
\beq
\kappa'=\frac{d+2}{2}\frac{nT}{m}\frac{1+\frac{3}{2}a_2}{\nu_{\kappa'}-\frac{3}{2}\zeta^{(0)}},
\label{n2.20}
\eeq
where
\beq
\nu_\kappa=\frac{\int d\mathbf{V}\,
\mathbf{S}(\mathbf{V})\cdot\mathcal{L}\boldsymbol{\mathcal{A}}(\mathbf{V})}
{\int d\mathbf{V}\,
\mathbf{S}(\mathbf{V})\cdot\boldsymbol{\mathcal{A}}(\mathbf{V})},\quad
\nu_{\kappa'}=\frac{\int d\mathbf{V}\,
\mathbf{S}(\mathbf{V})\cdot\mathcal{L}\boldsymbol{\mathcal{A}}'(\mathbf{V})}
{\int d\mathbf{V}\,
\mathbf{S}(\mathbf{V})\cdot\boldsymbol{\mathcal{A}}'(\mathbf{V})}.
\label{n2.21bis}
\eeq
The above expressions are formally \textit{exact}, but the
dependence of $\nu_\kappa$, $\nu_{\kappa'}$, $\zeta^{(0)}$, and
$a_2$ on $\alpha$ is unknown. While the two latter quantities
require the knowledge of the HCS distribution $f^{(0)}$, the
collision frequencies $\nu_\kappa$ and $\nu_{\kappa'}$ are given in
terms of the solutions of the two linear integral equations
(\ref{n2.11}) and (\ref{n2.13}). As said before, $\zeta^{(0)}$ and
$a_2$ can be well approximated by Eqs.\ (\ref{x2.29}) and
(\ref{n2.5}), respectively.

The symmetry properties of the vectorial functions
$\boldsymbol{\mathcal{A}}(\mathbf{V})$ and
$\boldsymbol{\mathcal{A}}'(\mathbf{V})$ suggest the following Sonine
expansion representations:
\beq
\left\{ \begin{array}{c} \boldsymbol{\mathcal{A}}(\mathbf{V})\\
\boldsymbol{\mathcal{A}}'(\mathbf{V})
\end{array}\right\}
=f_M(\mathbf{V})\sum_{k=1}^\infty
\left\{ \begin{array}{c} b_k\\
b_k'
\end{array}
\right\} L_k^{(d/2)}(c^2).
\label{n2.19bis}
\eeq
Making use of the orthogonality condition (\ref{5.4}), the
coefficients $b_k$ and $b_k'$ can be expressed as moments of
$\boldsymbol{\mathcal{A}}$ and $\boldsymbol{\mathcal{A}}'$,
respectively. In particular, the first coefficients $b_1$ and $b_1'$
are directly related to the thermal conductivities as
\beq
\left\{ \begin{array}{c} \kappa\\
\kappa'
\end{array}
\right\}= \frac{d+2}{4}n\lambda v_0\left\{ \begin{array}{c} b_1\\
b_1'
\end{array}
\right\}.
\label{n2.20bis}
\eeq

Unfortunately, when the expansions (\ref{n2.19bis}) are inserted
into Eqs.\ (\ref{n2.11}) and (\ref{n2.13}), one gets an infinite
hierarchy of  equations for the coefficients $b_k$ and $b_k'$, so
that $\kappa$ and $\kappa'$ cannot be obtained exactly. For
practical purposes, it is usual to truncate the Sonine expansions at
a given order $k$ and solve an approximate set of $k$  equations.
The case $k=1$ yields the so-called first Sonine approximation. In
this case, the collision frequencies defined in Eq.\
(\ref{n2.21bis}) are \cite{BC01}
\beqa
\nu_\kappa=\nu_{\kappa'}&=&\nu_0
\frac{1+\alpha}{d}\left[\frac{d-1}{2}+\frac{3}{16}(d+8)(1-\alpha)\right.
\nonumber\\
&&\left.+ \frac{4+5d-3(4-d)\alpha}{512}a_2\right].
\label{c11}
\eeqa
Insertion of Eqs.\ (\ref{x2.29}), (\ref{n2.5}), and (\ref{c11}) into
Eqs.\ (\ref{n2.19}) and (\ref{n2.20}) gives the transport
coefficients $\kappa$ and $\kappa'$ in the first Sonine
approximation. The coefficient $\mu$ is then obtained from Eq.\
(\ref{n2.17}). In the elastic case ($\alpha=1$), both $\kappa$ and
$\kappa'$ reduce to the thermal conductivity coefficient for elastic
hard spheres, which in the first Sonine approximation is given by
\beq
\kappa_0=\frac{d(d+2)}{2(d-1)}\frac{nT}{m\nu_0}.
\label{c12}
\eeq

\section{Homogeneous steady heat flow driven by a velocity-dependent external force}
\label{sec3}

\kk{ Apparently, the most direct way of measuring the NS velocity
distribution function (\ref{n2.8}) as well as its associated
transport coefficients $\kappa$ and $\kappa'$ would imply the
introduction of weak temperature and density gradients. However,
this gives rise to several technical problems. First, the coupling
between inelasticity and spatial gradients make it difficult to
extract the real NS contributions. Moreover, one must be able to
identify the bulk region, where the boundary effects are negligible.
In addition, the measured quantities are local and hence the
statistical errors become significant. Finally, it would be
difficult to disentangle the contributions to the heat flux coming
from the temperature and density gradients. This encourages the
search for alternative methods that avoid these difficulties.}

As said in the Introduction, we want to study a homogeneous
nonequilibrium steady state generated by the action of an
anisotropic  velocity-dependent external force
 of strength $\epsilon$, which induces a
heat flux in the absence of any thermal and density gradients. The
form of the force must be chosen under the condition that, in the
limit $\epsilon\to 0$, the deviation of the velocity distribution
function from that of the HCS is the same as the one produced by
real thermal and density gradients. In the latter situation, the
deviation from the HCS is measured by the NS functions
$\bm{\mathcal{A}}(\mathbf{V})$ and $\bm{\mathcal{A}}'(\mathbf{V})$,
as discussed in the preceding Section. Therefore, we have to deal
with two separate homogeneous steady Boltzmann equations, each one
reducing in the linear order in $\epsilon$ to Eqs.\ (\ref{n2.11})
and (\ref{n2.13}), respectively.

\subsection{Navier--Stokes function $\bm{\mathcal{A}}'(\mathbf{V})$}

Let us start with the function $\bm{\mathcal{A}}'(\mathbf{V})$
associated with the modified thermal conductivity $\kappa'$. Taking
into account the structure of Eq.\ (\ref{n2.13}), together with Eq.\
(\ref{n2.13bis}), we propose the following Boltzmann equation:
\beq
\frac{\partial}{\partial\mathbf{V}}\cdot\left[\mathbf{F}'(\mathbf{V})f(\mathbf{V})\right]=J[\mathbf{V}|f,f],
\label{13}
\eeq
where $\mathbf{F}'(\mathbf{V})$ is an external force (except for a
factor $m$). More specifically,
$\mathbf{F}'=\mathbf{F}'_\epsilon+\mathbf{F}_{\text{th}}$ is
decomposed into a heat-flux force
\beq
\mathbf{F}'_\epsilon(\mathbf{V})=\mathsf{G}'(\mathbf{V})\cdot\bm{\epsilon}
\label{11}
\eeq
and a thermostat force
\beq
\mathbf{F}_{\text{th}}(\mathbf{V})=\frac{\xi}{2}\mathbf{V}+\bm{\beta}.
\label{12}
\eeq
In Eq.\ (\ref{11}), the tensor $\mathsf{G}'(\mathbf{V})$ is given by
Eq.\ (\ref{n2.22}) and $\bm{\epsilon}$ is a constant vector which
mimics the effect of $\nabla \ln T$ (at $\nabla
nT^{1/2}=\mathbf{0}$) and whose magnitude $\epsilon$ measures the
strength of the force. The parameters $\xi$ and $\bm{\beta}$ in the
thermostat term are introduced to keep the total kinetic energy and
momentum constant, respectively.

Multiplying both sides of Eq.\ (\ref{13}) by $\mathbf{V}$,
integrating over velocity, and imposing a   vanishing total momentum
we get the following expression for the parameter $\bm{\beta}$:
\beq
\beta_i=\frac{1}{2mn}\Pi_{ij}\epsilon_j,
\label{13bis}
\eeq
where $\Pi_{ij}$ is defined by Eq.\ (\ref{2.8bis}).  Analogously,
the condition of constant total kinetic energy yields
\beq
\xi=\zeta+\frac{2}{dnT}\mathbf{q}\cdot\bm{\epsilon},
\label{14}
\eeq
where  the cooling rate $\zeta$ and the heat flux $\mathbf{q}$ are
defined by Eqs.\ (\ref{n1}) and (\ref{2.8}), respectively. Since the
direction of $\bm{\epsilon}$ can be chosen arbitrarily without loss
of generality, in what follows we take
$\bm{\epsilon}=\epsilon\widehat{\mathbf{x}}$. By symmetry reasons,
the only non-zero elements of  $\Pi_{ij}$ are $\Pi_{xx}$ and
$\Pi_{yy}=\Pi_{zz}=\cdots=\Pi_{dd}$. Analogously, $\mathbf{q}=q_x
\widehat{\mathbf{x}}$ and $\bm{\beta}=\beta_x \widehat{\mathbf{x}}$.
In addition, in the limit $\epsilon\to 0$, one has the leading
behaviors ${q}_x\sim \epsilon$ and $\Pi_{ij}\sim \epsilon^2$, so
that ${\beta}_x\sim\epsilon^3$ and $\xi-\zeta\sim\epsilon^2$.

In principle, Eq.\ (\ref{13}) is a highly nonlinear Boltzmann
equation since not only the collision term is quadratic in $f$, but
also the coefficients $\xi$ and $\bm{\beta}$ are functionals of $f$
through $\zeta$, $\Pi_{ij}$, and $\mathbf{q}$. The control parameter
in Eq.\ (\ref{13}) is the field strength $\epsilon$. Although the
problem of solving Eq.\ (\ref{13}) for finite $\epsilon$ is
interesting by itself \cite{L89,GS91}, here we focus on the regime
of small $\epsilon$. In that case, the solution to Eq.\ (\ref{13})
can be expanded in powers of $\epsilon$ as
\beq
f(\mathbf{V})=f^{(0)}(\mathbf{V})+f^{(1)}(\mathbf{V})+O(\epsilon^2).
\label{15}
\eeq
\kk{Note that the expansion (\ref{15}) does not need to be
convergent but only asymptotic, similarly to what happens in the CE
expansion.} Setting $\epsilon=0$ in Eq.\ (\ref{13}) and taking into
account that $\xi=\zeta^{(0)}+O(\epsilon^2)$, it is straightforward
to see that $f^{(0)}(\mathbf{V})$ verifies Eq.\ (\ref{n6}), i.e.,
the Boltzmann equation for the HCS. Next, to first order in
$\epsilon$, one has
\beq
\frac{\partial}{\partial\mathbf{V}}\cdot\left[\mathbf{F}'_\epsilon(\mathbf{V})f^{(0)}(\mathbf{V})+
\frac{1}{2}\zeta^{(0)} \mathbf{V}f^{(1)}(\mathbf{V})\right]=
-\mathcal{L}f^{(1)}(\mathbf{V}).
\label{n16}
\eeq
Therefore,
\beq
f^{(1)}(\mathbf{V})=\bm{\mathcal{A}}'(\mathbf{V})\cdot\bm{\epsilon},
\label{n3.1}
\eeq
where $\bm{\mathcal{A}}'(\mathbf{V})$ obeys the linear integral
equation (\ref{n2.13}) and use is made of Eq.\ (\ref{n2.13bis}).
This proves that the departure from the HCS distribution produced by
the external force $\mathbf{F}'(\mathbf{V})$ coincides to first
order in $\epsilon$ with the one produced by a real thermal
gradient. As a consequence, the modified thermal conductivity
$\kappa'$ can be obtained from the solution to Eq.\ (\ref{13})
through the linear response relation
\beq
\kappa'=-\frac{1}{T}\lim_{\epsilon\to 0}\frac{q_x}{\epsilon}.
\label{n3.2}
\eeq
Analogously, the coefficients $b_k'$ defined in Eq.\
(\ref{n2.19bis}) can be evaluated in terms of averages of velocity
polynomials:
\beq
b_k'=\frac{2\Gamma(1+d/2)k!}{\Gamma (k+1+d/2)} \lim_{\epsilon^*\to
0}\frac{\langle c_x L_k^{(d/2)}(c^2)\rangle}{\epsilon^*},
\label{5.7}
\eeq
where
\beq
\langle X(\mathbf{V})\rangle =\frac{1}{n}\int
d\mathbf{V}X(\mathbf{V})f(\mathbf{V})
\label{n3.3}
\eeq
and $\epsilon^* =\lambda\epsilon$ is the reduced field strength,
which plays the role of a Knudsen number. Since
$f^{(0)}(\mathbf{V})$ is an isotropic function, only
$f^{(1)}(\mathbf{V})$ contributes to the averages of Eq.\
(\ref{5.7}). Furthermore, it is proven in the Appendix  that the
coefficients $b_k'$ can alternatively be evaluated as
\beq
b_k'=\frac{2\Gamma(3/2)k!}{\Gamma (k+3/2)} \lim_{\epsilon^*\to
0}\frac{\langle c_x L_k^{(1/2)}(c_x^2)\rangle}{\epsilon^*}.
\label{5.7bis}
\eeq
While the averages in Eq.\ (\ref{5.7}) require the knowledge of the
full distribution $f^{(1)}(\mathbf{V})$, the averages in Eq.\
(\ref{5.7bis}) only require the knowledge of the marginal
distribution
\beq
g^{(1)}({V_x})=\int d\mathbf{V}_\perp \,f^{(1)}(\mathbf{V}),
\label{n3.4}
\eeq
where $\mathbf{V}_\perp\equiv \mathbf{V}-V_x\widehat{\mathbf{x}}$.
The equivalence between Eqs.\ (\ref{5.7}) and (\ref{5.7bis}) implies
that all the information contained in the full distribution
$f^{(1)}(\mathbf{V})$ is encapsulated in the marginal distribution
$g^{(1)}({V_x})$.

For sufficiently small values of $\epsilon$, Eq.\ (\ref{15}) can be
truncated, so that $f=f^{(0)}+f^{(1)}$. At the level of the
corresponding marginal distributions,
\beq
g({V_x})=g^{(0)}({V_x})+g^{(1)}({V_x}).
\label{15bis}
\eeq
By symmetry, $g^{(0)}({V_x})$ is an even function of $V_x$, while
$g^{(1)}({V_x})$ is an odd function. Consequently,
\beq
g^{(1)}({V_x})=\frac{1}{2}\left[g(V_x)-g(-V_x)\right].
\label{15bb}
\eeq
This relation is useful for extracting the first-order distribution
$g^{(1)}({V_x})$ from the complete distribution $g(V_x)$, provided
that $\epsilon$ is small enough to neglect nonlinear terms.

\subsection{Navier--Stokes function $\bm{\mathcal{A}}(\mathbf{V})$}
Comparison between the integral equations (\ref{n2.11}) and
(\ref{n2.13}) shows two differences. First, the inhomogeneous terms
are different, i.e., $\mathbf{A}(\mathbf{V})\neq
\mathbf{A}'(\mathbf{V})$. This means that the role of the force
(\ref{11}) is now played by the force
\beq
\mathbf{F}_\epsilon(\mathbf{V})=\mathsf{G}(\mathbf{V})\cdot\bm{\epsilon},
\label{11bis}
\eeq
where $\mathsf{G}(\mathbf{V})$ is given by Eq.\ (\ref{8}). The most
significant difference is the presence of the extra term
$-\frac{1}{2}\zeta^{(0)}\bm{\mathcal{A}}(\mathbf{V})$ on the
left-hand side of Eq.\ (\ref{n2.11}). This term cannot be accounted
for by an external force. Therefore, the appropriate homogeneous
steady Boltzmann equation in this case is
\beq
\frac{\partial}{\partial\mathbf{V}}\cdot\left[\mathbf{F}(\mathbf{V})f(\mathbf{V})\right]=J[\mathbf{V}|f,f]+\frac{1}{2}
\zeta\left(f-f^{(0)}\right),
\label{22}
\eeq
with $\mathbf{F}=\mathbf{F}_\epsilon+\mathbf{F}_{\text{th}}$, where
the heat-flux force is given by Eq.\ (\ref{11bis}) and the
thermostat force is again (\ref{12}). The condition of zero total
momentum yields
\beq
\beta_i=\frac{d-2}{2(d-1)mn}\Pi_{ij}\epsilon_j,
\label{23}
\eeq
while, by  the condition of constant energy, $\xi$ is still given by
Eq.\ (\ref{14}). Again, we can choose
$\bm{\epsilon}=\epsilon\widehat{\mathbf{x}}$. As before, in the
limit $\epsilon\to 0$, one has ${q}_x\sim \epsilon$, $\Pi_{ij}\sim
\epsilon^2$, ${\beta}_x\sim\epsilon^3$, and
$\xi-\zeta\sim\epsilon^2$. Inserting the expansion (\ref{15}) into
Eq.\ (\ref{22}), and following the same steps as in the preceding
subsection, one sees that $f^{(0)}$ is the HCS distribution, Eq.\
(\ref{n6}), and
\beq
f^{(1)}(\mathbf{V})=\bm{\mathcal{A}}(\mathbf{V})\cdot\bm{\epsilon},
\label{n3.1bis}
\eeq
where $\bm{\mathcal{A}}(\mathbf{V})$ obeys the linear integral
equation (\ref{n2.11}). Thus, the thermal conductivity $\kappa$ can
be obtained from the solution to Eq.\ (\ref{22}) through a linear
response relation similar to Eq.\ (\ref{n3.2}). Analogously, the
Sonine coefficients $b_k$ can be evaluated from expressions similar
to Eqs.\ (\ref{5.7}) and (\ref{5.7bis}).

The main peculiarity of the Boltzmann equation  (\ref{22}) lies in
the presence of the anomalous term
$\frac{1}{2}\zeta\left(f-f^{(0)}\right)$ on the right-hand side.
This term represents a \textit{stochastic} clonation-annihilation
process. According to it, a particle of a given velocity
$\mathbf{V}$ has a probability per unit time $\frac{1}{2}\zeta$ of
being ``cloned'' and a probability per unit time $\frac{1}{2}\zeta
f^{(0)}(\mathbf{V})/f(\mathbf{V})$ of being ``annihilated.'' Since
$f$ and $f^{(0)}$ are normalized to the same number density, flow
velocity, and temperature, the clonation-annihilation process does
not change the total number of particles, momentum, and energy.
However, this stochastic process tends to increase the departure of
the distribution function from that of the HCS. It is worth
remarking that, in the homogeneous steady Boltzmann equation
associated with the shear viscosity \cite{MSG05}, a similar
stochastic process is also present, but with the opposite sign, so
that the stochastic term tends to decrease the departure from of the
HCS. The presence or absence of this type of stochastic term has a
counterpart in the GK relations. While the time correlation function
is multiplied  by  an exponentially growing factor  in the case of
$\kappa$, there is an exponentially decaying factor for $\eta$, and
there is no factor in the case of $\kappa'$ \cite{DB02}.

The presence of the non-standard term
$\frac{1}{2}\zeta\left(f-f^{(0)}\right)$ makes it difficult, \kk{but
not impossible,} to implement the Boltzmann equation (\ref{22}) by
the DSMC method. Thus, for the sake of simplicity, henceforth we
will focus on the Boltzmann equation (\ref{13}) to determine the NS
distribution function $\bm{\mathcal{A}}(\mathbf{V})$ and its
associated transport coefficient $\kappa'$.

\kk{Before closing this Section, it is worth mentioning that the
heat-flux forces (\ref{11}) and (\ref{11bis}) are not
straightforward extensions of the one proposed by Evans \cite{E82}
and Gillan and Dixon \cite{GD83} for normal fluids. The latter is
$\mathbf{F}_\epsilon^{\text{el}}(\mathbf{V})=\mathsf{G}^{\text{el}}(\mathbf{V})\cdot\bm{\epsilon}$,
where
\beq
{G}_{ij}^{\text{el}}(\mathbf{V})=\left(\frac{dT}{2m}-\frac{V^2}{2}\right)\delta_{ij}.
\label{EGD}
\eeq
As a consequence, $\mathbf{F}_\epsilon^{\text{el}}(\mathbf{V})\|
\bm{\epsilon}$, while $\mathbf{F}_\epsilon(\mathbf{V})$  and
$\mathbf{F}_\epsilon'(\mathbf{V})$ are not parallel to
$\bm{\epsilon}$. For $\alpha<1$,
$\partial_{\mathbf{V}}\cdot\left[\mathsf{G}^{\text{el}}(\mathbf{V})f^{(0)}(\mathbf{V})\right]\neq
\partial_{\mathbf{V}}\cdot\left[\mathsf{G}(\mathbf{V})f^{(0)}(\mathbf{V})\right]\neq
\partial_{\mathbf{V}}\cdot\left[\mathsf{G}'(\mathbf{V})f^{(0)}(\mathbf{V})\right]$. However, in the elastic case ($\alpha=1$),
$f^{(0)}(\mathbf{V})=f_M(\mathbf{V})$, so that
\beqa
\frac{\partial}{\partial{\mathbf{V}}}\cdot\left[\mathsf{G}^{\text{el}}(\mathbf{V})f_M(\mathbf{V})\right]&=&
\frac{\partial}{\partial{\mathbf{V}}}\cdot\left[\mathsf{G}(\mathbf{V})f_M(\mathbf{V})\right]\nn
&=&
\frac{\partial}{\partial{\mathbf{V}}}\cdot\left[\mathsf{G}'(\mathbf{V})f_M(\mathbf{V})\right]\nn
&=&T^{-1}\mathbf{S}(\mathbf{V})f_M(\mathbf{V}),
\label{EGD2}
\eeqa
} where $\mathbf{S}(\mathbf{V})$ is given by Eq.\ (\ref{n2.16bis}).

\section{Results}
\label{sec4}
In this Section we present results obtained by numerically solving
the Boltzmann equation (\ref{13}) in the three-dimensional case
($d=3$) by means of the DSMC method \cite{DSMC}. \kk{It has  been
rigorously proven that the DSMC method produces a solution to the
Boltzmann equation in the limit of vanishing discretization and
stochastic errors \cite{W92}.} Details of the method can be found
elsewhere \cite{DSMC}, so that here we only point out some specific
aspects in our simulations. The system consists of $N$ simulated
particles, whose velocities are updated from time $t$ to time
$t+\delta t$ due to (i) the collisions and (ii) the action of the
external force $\mathbf{F}'(\mathbf{V})$. The collisional stage (i)
proceeds as usual \cite{DSMC}, except that the collisions are
inelastic. The streaming stage (ii) proceeds in two steps:
\begin{enumerate}
\item
\kk{Update the velocity of every particle $\ell=1,\ldots,N$ as
follows:
\beq
\mathbf{V}'_\ell(t+\delta
t)=\mathbf{V}_\ell(t)-\frac{1}{2}\mathbf{V}_\ell(t)\mathbf{V}_\ell(t)\cdot
\bm{\epsilon} \delta t.
\label{16}
\eeq
This only accounts for the velocity-dependent part of the heat-flux
term $\mathbf{F}_\epsilon'(\mathbf{V})$, as can be seen from Eq.\
(\ref{n2.22}).
\item
Compute the mean velocity $\mathbf{u}'(t+\delta
t)=N^{-1}\sum_{\ell=1}^N \mathbf{V}'_\ell(t+\delta t)$ and the
kinetic energy per particle $K'(t+\delta
t)=N^{-1}(m/2)\sum_{\ell=1}^N [\mathbf{V}'_\ell(t+\delta
t)-\mathbf{u}'(t+\delta t)]^2$. In general, $\mathbf{u}'(t+\delta
t)\neq \mathbf{0}$ because of the step (\ref{16}) and $K'(t+\delta
t)\neq K_0$, where $K_0$ is the prescribed initial kinetic energy
per particle, because of the collisional stage and also because of
the step (\ref{16}). Next, update again the velocity of every
particle as
\beq
 \mathbf{V}_\ell(t+\delta t)=
\sqrt{\frac{K_0}{K'(t+\delta t)}}\left[\mathbf{V}_\ell'(t+\delta
t)-\mathbf{u}'(t+\delta t)\right].
\label{17}
\eeq
}This change guarantees that the total momentum is restored to zero
and the kinetic energy to its initial value, before proceeding again
to the collisional stage. Equation (\ref{17}) incorporates the
effect of $\mathbf{F}_{\text{th}}$, as well as of the
velocity-independent part of $\mathbf{F}_\epsilon'$.
\end{enumerate}

Once a steady state is reached, the most  relevant quantities can be
computed. This includes the heat flux $q_x$, the averages $\langle
c_x L_k^{(3/2)}(c^2)\rangle$ and $\langle c_x
L_k^{(1/2)}(c_x^2)\rangle$ with $k=1,2,3$, and the marginal velocity
distribution function $g(V_x)$. Note that $\kappa'$ and $b_1'$ are
essentially the same quantity, as shown by Eq.\ (\ref{n2.20bis}). By
assuming that the value of the reduced strength $\epsilon^* $  is
small enough to probe the linear regime, the modified thermal
conductivity $\kappa'$ and the Sonine coefficients $b_1'$, $b_2'$,
and $b_3'$ are obtained from Eqs.\ (\ref{n3.2}), (\ref{5.7}), and
(\ref{5.7bis}). In addition, the marginal NS distribution function
$g^{(1)}(V_x)$ is obtained from $g(V_x)$ by applying Eq.\
(\ref{15bb}).

The number of simulated particles is $N=2\times 10^5$ and the time
step is $\delta t=0.003 \tau$, where $\tau=\lambda/v_0$ is the mean
free time and $\lambda=1/\sqrt{2}\pi n\sigma^2$. To improve
statistics, the results are averaged over 200 independent
realizations. The data are further averaged over time in the steady
regime. The range of inelasticities analyzed is $0.3\leq\alpha\leq
1$ and in all the cases the initial state is a Gaussian
distribution.

\begin{figure}[htb]
\includegraphics[width=\columnwidth]{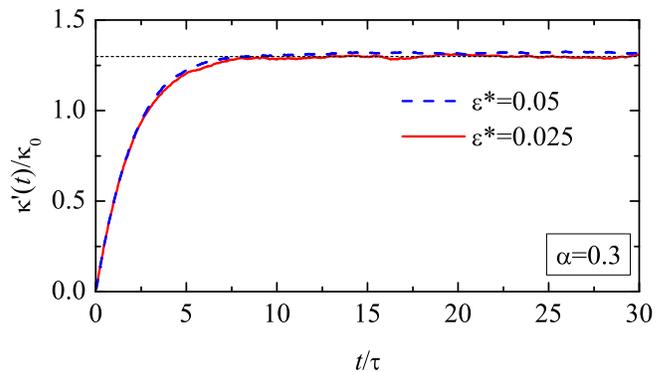}
\caption{(Color online) Plot of the  time-dependent (modified)
thermal conductivity versus time for $\alpha=0.3$ and $\epsilon^*=
0.05$ (dashed line) and $\epsilon^*= 0.025$ (solid line). The
horizontal dotted line indicates the steady-state value averaged
over time  in the case $\epsilon^*= 0.025$.
\label{fig1}}
\end{figure}

We first analyze the evolution toward the steady state. In the
transient regime one can measure effective time-dependent values of
$\kappa'$ and $b_k'$. Figure \ref{fig1} shows the ratio
$\kappa'(t)/\kappa_0$, where $\kappa_0$ is defined by Eq.\
(\ref{c12}), versus $t/\tau$ for $\alpha=0.3$ and two different
values of the reduced field strength, namely $\epsilon^*=0.05$  and
$\epsilon^*=0.025$. Even in this highly inelastic case, we observe
that a steady-state value of $\kappa'$ is reached after about ten
collisions per particle. Also, the figure clearly indicates that the
transient and the stationary results are very weakly dependent of
$\epsilon^*$, i.e., the heat flux $q_x$ is practically linear in
$\epsilon^*$. We have observed a similar behavior for  other values
of $\alpha$. Thus,  we take the value $\epsilon^*=0.025$ in the
remainder of the figures. An even smaller value of $\epsilon^*$
would decrease the signal-to-noise ratio without affecting the
results.

\begin{figure}[htb]
\includegraphics[width=\columnwidth]{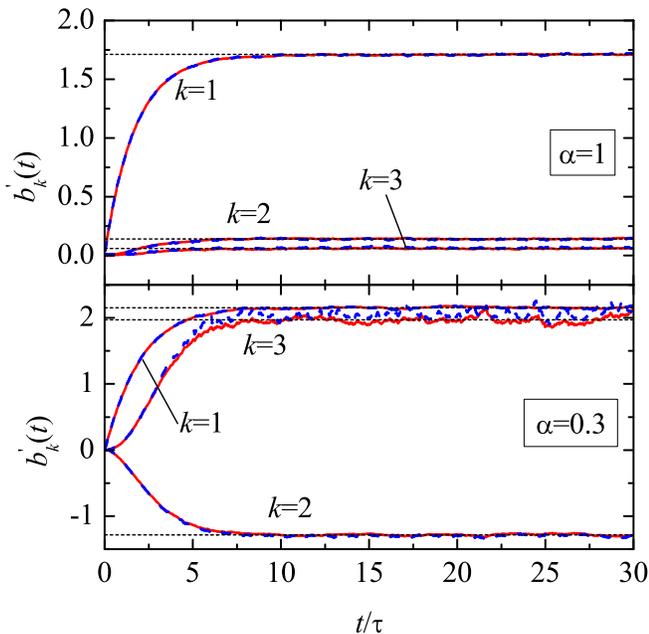}
\caption{(Color online) Plot of the  time-dependent Sonine
coefficients $b_k'$ ($k=1,2,3$) versus time for $\alpha=1$ (top
panel) and $\alpha=0.3$ (bottom panel). The solid and dahed lines
correspond to data obtained from $\langle c_x
L_k^{(3/2)}(c^2)\rangle$ and from $\langle c_x
L_k^{(1/2)}(c_x^2)\rangle$, respectively. Note that they are
distinguishable only in the case of $b_3'$ for $\alpha=0.3$. The
horizontal dotted lines indicate the steady-state values.
\label{fig2}}
\end{figure}

An additional test of the linear regime is provided by Fig.\
\ref{fig2}, which shows the time evolution of the Sonine
coefficients $b_1'$, $b_2'$, and $b_3'$ for $\alpha=1$ and
$\alpha=0.3$. These coefficients have been evaluated from Eq.\
(\ref{5.7}) and also from Eq.\ (\ref{5.7bis}). As proven in the
Appendix,  both methods must lead to identical results, provided
that the system is in the linear regime. Figure \ref{fig2} shows
that the simulation data obtained from  $\langle c_x
L_k^{(3/2)}(c^2)\rangle$ are practically indistinguishable from
those obtained from $\langle c_x L_k^{(1/2)}(c_x^2)\rangle$, thus
confirming that the value $\epsilon^*=0.025$ is small enough to
assume that the system is actually in the linear regime. Given that
the fluctuations are smaller when the Sonine coefficients are
computed from $\langle c_x L_k^{(3/2)}(c^2)\rangle$ than when they
are computed from $\langle c_x L_k^{(1/2)}(c_x^2)\rangle$,
henceforth we choose the former method. We also see in Fig.\
\ref{fig2} that the characteristic relaxation time ($t/\tau \sim
10$) is practically independent of the degree of inelasticity as
well as of the order of the Sonine coefficient considered.

\begin{figure}[htb]
\includegraphics[width=\columnwidth]{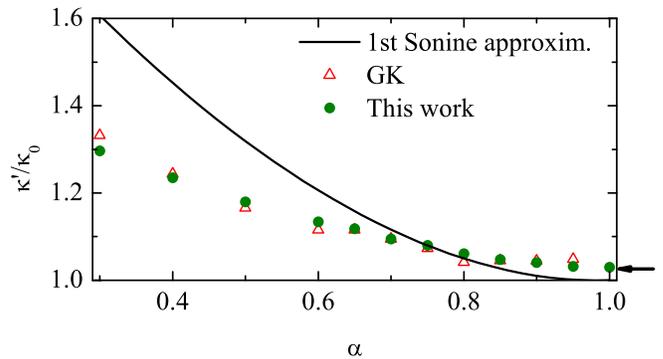}
\caption{(Color online) Plot of the  modified thermal conductivity
coefficient $\kappa'$ versus the coefficient of normal restitution
$\alpha$. The circles refer to the simulation data obtained by the
method  described in this paper, while the triangles refer to the
simulation data obtained in Ref.\ \protect\cite{BRMMG05} from the GK
relations. \kk{We have checked that the error bars of our simulation
data are smaller than the size of the symbols.} The solid line is
the prediction given by the first Sonine approximation. \kk{The
arrow indicates the value $\kappa/\kappa_0=1.025218$ corresponding
to the elastic limit when higher order terms in the Sonine expansion
are taken into account.}
\label{fig3}}
\end{figure}
Now we focus on the dependence of the steady-state quantities on
dissipation. Figure  \ref{fig3} shows $\kappa'/\kappa_0$ as a
function of the coefficient of restitution $\alpha$. In addition to
our simulation data, we have included the values of $\kappa'$
obtained from the simulation data of $\kappa$ and $\mu$ reported by
Brey et al.\ \cite{BRMMG05}, as well as the first Sonine
approximation given by Eqs.\ (\ref{n2.20}) and (\ref{c11}). It must
be emphasized that the methods used here and in Ref.\ \cite{BRMMG05}
are quite different. In the latter method,  two-time correlation
functions of certain quantities are evaluated in the HCS and
subsequently those correlation functions are integrated in time to
get the transport coefficients. On the other hand, in our method the
HCS is slightly perturbed by a homogeneous external forcing and the
transport coefficient is determined from a standard one-time average
(namely, the heat flux), by assuming  linear response. While the
first Sonine prediction does a good job for $\alpha\gtrsim 0.7$, it
dramatically overestimates the thermal conductivity for higher
inelasticity. It is important to remark that the excellent agreement
found between both simulation methods strongly supports the
conclusion that the discrepancies between the first Sonine
approximation and simulations are not due to the presence of
velocity correlations beyond the Boltzmann description for strong
dissipation. Instead, those discrepancies are \kk{essentially} due
to the inaccuracy of the first Sonine approximation for
$\alpha\lesssim 0.7$. \kk{In the elastic limit, the numerical
results show that the first Sonine approximation slightly
underestimates the thermal conductivity. This is known to be
corrected if higher orders in the Sonine polynomial expansion are
taken into account \cite{CC70}, which yields
$\kappa/\kappa_0=1.025218$ \cite{Gallis}. This value is indicated by
an arrow in Fig.\ \ref{fig3} and agrees with our simulation result
for $\alpha=1$.}

\begin{figure}[htb]
\includegraphics[width=\columnwidth]{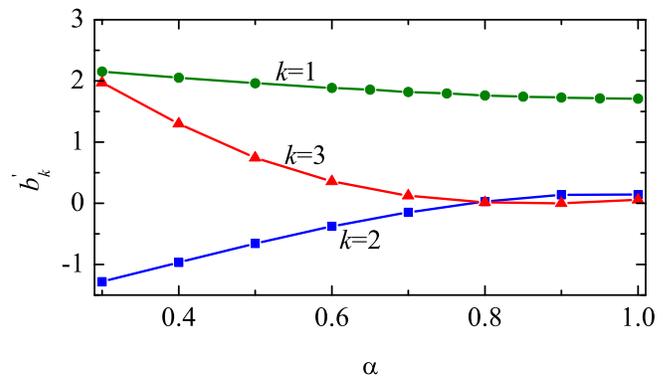}
\caption{(Color online) Plot of the  Sonine coefficients $b_1'$
(circles), $b_2'$ (squares), and $b_3'$ (triangles) versus the
coefficient of normal restitution $\alpha$. \kk{The error bars are
smaller than the size of the symbols.} The lines are guides to the
eye.
\label{fig4}}
\end{figure}
The above conclusion \kk{about the strong inaccuracy of the first
Sonine approximation for $\alpha\lesssim 0.7$} is further confirmed
by Fig.\ \ref{fig4}, which shows the $\alpha$-dependence of the
three first Sonine coefficients. We observe that $b_1'$ is much less
sensitive to dissipation than $b_2'$ and $b_3'$. For not too large
inelasticity ($\alpha\gtrsim 0.7$), the second and third Sonine
coefficients are much smaller than $b_1'$ and, consequently, one may
expect that the series (\ref{n2.19bis}) truncated after $k=1$ is
sufficiently close to the true distribution $\bm{\mathcal{A}}'$, at
least in the region of thermal velocities relevant for the
evaluation of the heat flux. On the other hand, as dissipation
increases beyond $\alpha\approx 0.7$, the  magnitude of the
coefficients $b_2'$ and $b_3'$ grow rapidly, becoming comparable to
$b_1'$. This implies that contributions to $\bm{\mathcal{A}}'$
beyond the first Sonine term cannot be neglected. In principle, the
discrepancies of the first Sonine approximation could be remedied by
considering the second Sonine approximation, i.e., by truncating the
expansion (\ref{n2.19bis}) after $k=2$. However, according to Fig.\
\ref{fig4},  there does not exist a range of values of $\alpha$
where the magnitude of $b_3'$ is much smaller than that of $b_2'$,
so that the second Sonine approximation possibly would not be
sufficient.

It is worthwhile remarking that the characteristic value of $\alpha$
approximately below which the first Sonine approximation begins to
deviate significantly from the simulation data is also the value
below which the fourth cumulant $a_2$ of the HCS \kk{starts to grow
\cite{MS00,CDPT03,BCRM99,BP06}.} This means that there seems to be a
close relationship between the deviation of the HCS distribution
$f^{(0)}$ from its Gaussian form and the deviation of the NS
distribution $f^{(1)}$ from its first Sonine approximation. Given
that the weight function in the Sonine expansion is the Gaussian, it
seems natural to conjecture that a better estimate of $f^{(1)}$ can
be obtained by replacing the Gaussian weight function by the HCS
distribution $f^{(0)}$ \cite{L05,GS06}.

One of the advantages of the simulation method devised in this paper
is that it allows to measure not only the transport coefficient
$\kappa'$ or higher velocity moments (such as those associated with
$b_2'$ and $b_3'$), but also the NS velocity distribution function
itself. As said in Sec.\ \ref{sec3}, all the information contained
in the full distribution $f^{(1)}(\mathbf{V})$ is present in the
marginal distribution function $g^{(1)}({V}_x)$ defined by Eq.\
(\ref{n3.4}). Thus, in what follows, we restrict ourselves to this
marginal distribution function, which can be computed more
efficiently than $f^{(1)}(\mathbf{V})$ in the simulations. In order
to analyze $g^{(1)}({V}_x)$, it is convenient to write it in the
form
\beq
g^{(1)}({V}_x)=n
v_0^{-1}\pi^{-1/2}e^{-c_x^2}c_x\varphi(c_x^2)\epsilon^*,
\label{4.1}
\eeq
where $\varphi(c_x^2)$ is a dimensionless isotropic function,
independent of $\epsilon^*$. Its Sonine expansion is
\beq
\varphi(c_x^2)=\sum_{k=1}^\infty b_k' L_k^{(1/2)}(c_x^2),
\label{4.2}
\eeq
where, as proven in the Appendix, the coefficients $b_k'$ are the
same as in the Sonine expansion (\ref{n2.19bis}). For the sake of
comparison, it is convenient to introduce the \textit{truncated}
series
\beq
\varphi_p(c_x^2)=\sum_{k=1}^p b_k' L_k^{(1/2)}(c_x^2).
\label{4.3}
\eeq
If the first Sonine approximation is reliable, this means that
$\varphi(c_x^2)\simeq \varphi_1(c_x^2) \equiv
b_1'(\frac{3}{2}-c_x^2)$ in the region of thermal velocities (say
$c_x^2\lesssim 10$). In addition, one would expect that an even
better approximation would be obtained with $\varphi_2(c_x^2)$ and
$\varphi_3(c_x^2)$.

\begin{figure}[htb]
\includegraphics[width=\columnwidth]{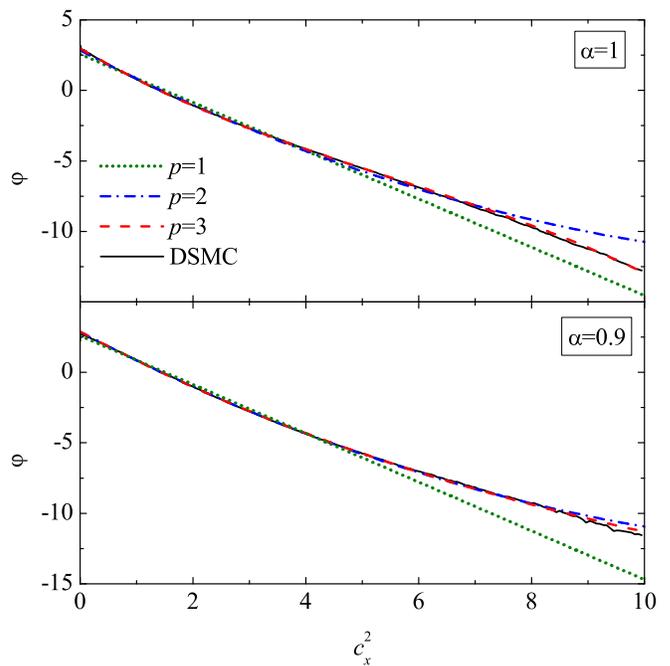}
\caption{(Color online) Plot of the  distribution function
$\varphi(c_x^2)$ (solid lines) for $\alpha=1$ (top panel) and
$\alpha=0.9$ (bottom panel). The dotted, dotted-dashed, and dashed
lines correspond to the truncated Sonine expansions
$\varphi_p(c_x^2)$ with $p=1$, 2, and 3, respectively.
\label{fig5}}
\end{figure}

\begin{figure}[htb]
\includegraphics[width=\columnwidth]{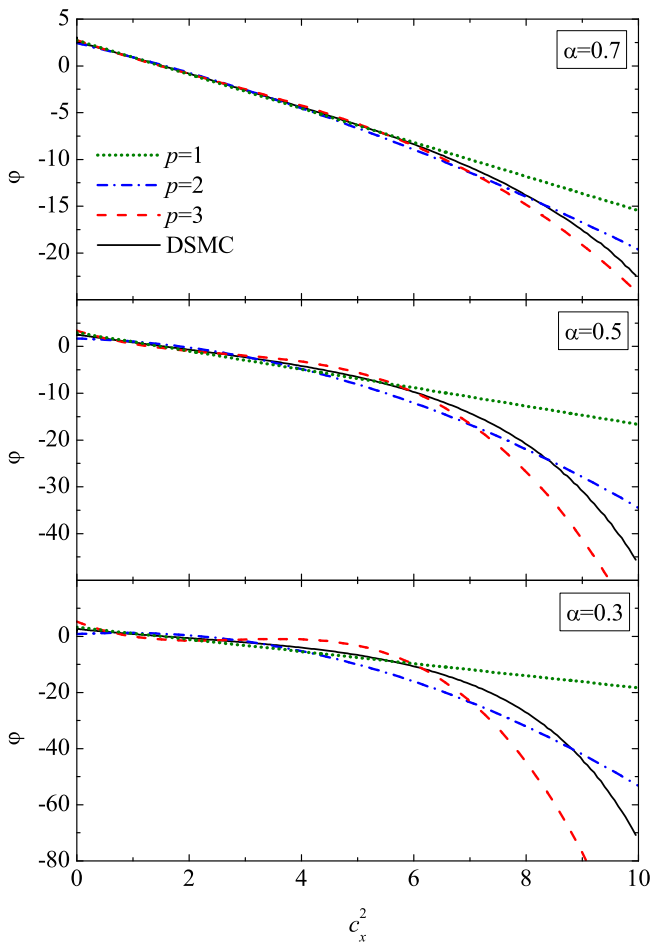}
\caption{(Color online) Plot of the  distribution function
$\varphi(c_x^2)$ (solid lines) for $\alpha=0.7$ (top panel),
$\alpha=0.5$ (middle panel),  and $\alpha=0.3$ (bottom panel). The
dotted, dotted-dashed, and dashed lines correspond to the truncated
Sonine expansions $\varphi_p(c_x^2)$ with $p=1$, 2, and 3,
respectively.
\label{fig6}}
\end{figure}

The function $\varphi(c_x^2)$ is plotted in Figs.\ \ref{fig5} (for
$\alpha=1$ and $\alpha=0.9$) and \ref{fig6} (for $\alpha=0.7$,
$\alpha=0.5$, and $\alpha=0.3$). The first three truncated
polynomials $\varphi_p(c_x^2)$, $p=1,2,3$, obtained by using the
simulation values of $b_p'$, are also plotted. In Fig.\ \ref{fig5}
we observe that the first Sonine polynomial $\varphi_1(c_x^2)$
captures reasonably well the behavior of the true distribution
$\varphi(c_x^2)$, although it underestimates the latter for
$c_x^2\gtrsim 4$. The addition of the second and third Sonine
polynomials significantly improves the agreement with the true
distribution. In fact, $\varphi_3$ is practically indistinguishable
from $\varphi$.
As the inelasticity increases, so does the deviation of $\varphi$
from the first Sonine polynomial, as shown in Fig.\ \ref{fig6}. In
contrast to the situation in Fig.\ \ref{fig5}, $\varphi_1$
overestimates the function $\varphi$ for large velocities. Moreover,
the second and third Sonine truncated expansions do not clearly
improve the agreement, especially for $\alpha=0.5$ and $\alpha=0.3$.

The comparison carried out in Figs.\ \ref{fig5} and \ref{fig6}
confirms that, if $\alpha\lesssim 0.7$, the velocity distribution
function is not sufficiently well described by the first term in the
Sonine expansion  and, consequently, the value of $b_1'$ (and hence
of $\kappa'$) estimated from the first Sonine approximation is not
accurate. It is also interesting to note that when $\varphi_1$
underestimates (overestimates) the function $\varphi$ for large
velocities, $\kappa'$ tends to be underestimated (overestimated) by
the first Sonine approximation. In view of the panels corresponding
to $\alpha=0.5$ and $\alpha=0.3$ in Fig.\ \ref{fig6}, it is highly
questionable that the second or third Sonine approximations could
provide accurate estimates for $\kappa'$. An alternative route has
been recently proposed \cite{GS06}.

\begin{figure}[htb]
\includegraphics[width=\columnwidth]{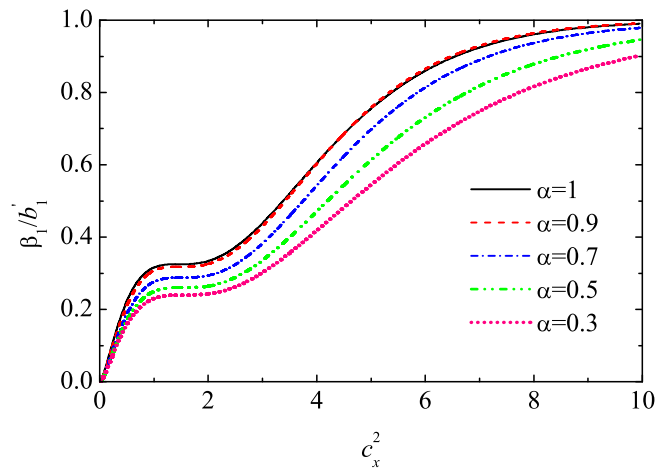}
\caption{(Color online) Plot of the  ratio $\beta_1(c_x^2)/b_1'$ for
$\alpha=1$, 0.9, 0.7, 0.5, and 0.3.
\label{fig7}}
\end{figure}
Finally, we address the question on the velocity range relevant to
the evaluation of $\kappa'\propto b_1'$. According to Figs.\
\ref{fig5} and \ref{fig6}, if that range were, for instance, $0\leq
c_x^2\lesssim 6$, then the first Sonine estimate would be reliable
even for $\alpha=0.3$. Since this is not the case, it is obvious
that the population of particles with $c_x^2\gtrsim 6$ must have a
significant influence on the heat flux for large inelasticity. To
analyze this point with more detail, let us introduce the function
\beq
\beta_1(c_x^2)=\frac{8}{3\pi^{1/2}}\int_0^{c_x^2}du_x\,
u_x^2\left(\frac{3}{2}-u_x^2\right)e^{-u_x^2}\varphi(u_x^2).
\label{4.4}
\eeq
This function represents the contribution to $b_1'$ coming from
particles whose $x$-component of the (reduced) velocity is smaller
than $|c_x|$, so that $\lim_{c_x^2\to\infty} \beta_1(c_x^2)=b_1'$.
The ratio $\beta_1(c_x^2)/b_1'$ is plotted in Fig.\ \ref{fig7} for
the same values of $\alpha$ as in Figs.\ \ref{fig5} and \ref{fig6}.
We see that the larger the inelasticity, the wider the range of
velocities needed to faithfully evaluate $b_1'$. For instance,
particles with $0\leq c_x^2\leq 6$ contribute to 87\% of the heat
flux in the elastic case, while that percentage reduces to 66\% in
the case of $\alpha=0.3$.

\section{Conclusions\label{sec5}}

The CE solution of the inelastic Boltzmann equation provides
expressions for the NS transport coefficients in terms of the
solutions of linear integral equations \cite{BDKS98}. Alternatively,
these expressions can be proven to be equivalent to GK relations
\cite{DB02}. In either case, in order to obtain the explicit
dependence of the transport coefficients on the coefficient of
restitution, one needs to make use of certain approximations. As in
the case of elastic collisions, the standard approach is to expand
the unknown NS velocity distribution function in Sonine polynomials
and truncate it at a given order. Of course, the practical
difficulties increase considerably with the number of retained
polynomials, so the first Sonine approximation is usually chosen. To
check the reliability of the first Sonine approximation one needs to
resort to comparison with DSMC computer simulations of the Boltzmann
equation. In principle, there are two strategies to measure the
transport coefficients via computer simulations in
\textit{homogeneous} states. One consists of measuring the
appropriate time correlation functions in the HCS and then carry out
an integration over time by applying the GK relations \cite{DB02}.
In the other possible strategy, one weakly disturbs the granular gas
from the HCS by the action of a conveniently chosen homogeneous,
anisotropic external force that produces the same effect as a
hydrodynamic gradient; in this way, the simulations provide in the
linear response regime the transport coefficients as well as the NS
velocity distribution functions. The first method has been used to
get the shear viscosity $\eta$ and the two transport coefficients
($\kappa$ and $\mu$) associated with the heat flux
\cite{BRM04,BRMMG05}, while so far the second method has only been
applied  to $\eta$ \cite{MSG05}.

The first Sonine approximation for $\eta$ is seen to compare quite
well with DSMC simulations, even for strong dissipation
\cite{BRM04,BRMMG05,MSG05}. However, the corresponding
approximations for $\kappa$ and $\mu$ appreciably differ from GK
simulations for high inelasticity ($\alpha\lesssim 0.7$). \kk{One
could argue} that these discrepancies are a reflection of the
velocity correlations appearing in the HCS, even in the low-density
limit. In that scenario, the correlation functions computed in DSMC
simulations would incorporate effects not accounted for by the
Boltzmann equation.

The primary goal of this paper has been to investigate whether the
discrepancies observed between the first Sonine estimates and the GK
data are due to effects beyond the Boltzmann framework or are simply
due to the limitations of the first Sonine approach. To that end, we
have followed the second strategy mentioned above, namely, the one
based on the action of a homogeneous external force. First, we have
derived the explicit forms of the respective velocity-dependent
external forces which yield, to first order, the NS velocity
distribution functions associated with the standard thermal
conductivity $\kappa$ and the modified thermal conductivity
$\kappa'\equiv \kappa-n\mu/2T$. In the case of $\kappa$, the
external force must be complemented by a stochastic term describing
a clonation-annihilation process. Since the presence of this latter
term complicates the simulation method, in this paper we have
focused on the homogeneous Boltzmann equation corresponding to the
coefficient $\kappa'$. Our DSMC results show an excellent
consistency with those obtained in Ref.\ \cite{BRMMG05} by the GK
formalism. Since our method is entirely tied to the Boltzmann
equation, we conclude that the deviations of the simulation data
from the first Sonine approximation are mainly due to the inaccuracy
of the latter. This conclusion has been further supported by an
analysis of the first three Sonine coefficients and of the NS
velocity distribution function itself. While the second and third
Sonine coefficients are practically negligible for $\alpha\gtrsim
0.7$, they rapidly increase in magnitude as the inelasticity
increases, becoming comparable to the first Sonine coefficient.
 With respect to the distribution function, we
have observed that, for strong dissipation, it is not well captured
by the first Sonine polynomial in the whole velocity region relevant
to the computation of the transport coefficient $\kappa'$.

The simulation results reported in this paper indicate that the
second or third Sonine approximations are not expected to improve
significantly the quality of the first Sonine estimate, especially
considering the technical difficulties associated with the method.
An alternative avenue under the form of a modified first Sonine
approximation is explored elsewhere \cite{GS06}.

\begin{acknowledgments}
This research  has been supported by the Ministerio de Educaci\'on y
Ciencia (Spain) through grants Nos.\ ESP2003-02859 (J.M.M.) and
FIS2004-01399 (A.S. and V.G.), partially financed by FEDER funds.
\end{acknowledgments}

\appendix*

\section{Proof of the equivalence between Eqs.\ (\protect\ref{5.7}) and
\lowercase{(\protect\ref{5.7bis})}}
\label{appB}

The full NS velocity distribution function (\ref{n3.1})
 can be written as
\beq
f^{(1)}(\mathbf{V})=n
v_0^{-d}\pi^{-d/2}e^{-c^2}c_x\Phi(c^2)\epsilon^*,
\label{B4.1}
\eeq
where $\Phi(c^2)$ is a dimensionless isotropic function.
 This function  can
be expanded in Laguerre (or Sonine) polynomials as
\beq
\Phi(c^2)= \sum_{k=1}^\infty b_k' L_k^{(d/2)}(c^2),
\label{5.2}
\eeq
in agreement with Eq.\ (\ref{n2.19bis}). The orthogonality condition
(\ref{5.4}) yields
\beq
b_k'=\frac{2\Gamma(1+d/2)k!}{\Gamma (k+1+d/2)}\pi^{-d/2} \int
d\mathbf{c}\, c_x^2 e^{-c^2}L_k^{(d/2)}(c^2)\Phi(c^2).
\label{5.5}
\eeq
This expression is equivalent to Eq.\ (\ref{5.7}).

Let us now consider  the marginal distribution $g^{(1)}(V_x)$
defined by Eq.\ (\ref{n3.4}). Similarly to Eq.\ (\ref{B4.1}),
$g^{(1)}(V_x)$ can be expressed as Eq.\ (\ref{4.1}). Thus, the
function $\varphi(c_x^2)$ can be obtained from $\Phi(c^2)$ as
\beq
\varphi(c_x^2)=\pi^{-(d-1)/2}\int d\mathbf{c}_\perp\,
e^{-c_\perp^2}\Phi(c^2).
\label{5.12}
\eeq
Inserting the expansion (\ref{5.2}) we get
\beq
\varphi(c_x)=\sum_{k=1}^\infty b_k' F_k(c_x^2),
\label{5.15}
\eeq
where we have called
\beq
F_k(c_x^2)\equiv \pi^{-(d-1)/2}\int d\mathbf{c}_\perp\,
e^{-c^2_\perp} L_k^{(d/2)}(c^2).
\label{5.16}
\eeq
Now we make use of the mathematical property \cite{AS72}
\beq
L_{k}^{(p+q+1)}(x+y)=\sum_{\ell=0}^k
L_\ell^{(p)}(x)L_{k-\ell}^{(q)}(y)
\label{5.17}
\eeq
and take $p=1/2$, $q=(d-3)/2$, $x=c_x^2$, and $y=c_\perp^2$. Thus,
\beq
F_k(c_x^2)=\pi^{-(d-1)/2}\sum_{\ell=0}^k L_\ell^{(1/2)}(c_x^2)\int
d\mathbf{c}_\perp\, e^{-c^2_\perp}
L_{k-\ell}^{(\frac{d-3}{2})}(c_\perp^2).
\label{B1}
\eeq
The integral can be computed as
\beqa
&& \pi^{-(d-1)/2}\int d\mathbf{c}_\perp\, e^{-c^2_\perp}
L_{k-\ell}^{(\frac{d-3}{2})}(c_\perp^2)\nn &&
=\frac{1}{\Gamma((d-1)/2)}\int_0^\infty dy\,
y^{(d-3)/2}L_{k-\ell}^{(\frac{d-3}{2})}(y)\nn && =\delta_{k,\ell},
\label{B2}
\eeqa
where in the last step we have made use of the orthogonality
relation of the generalized Laguerre polynomials. Therefore,
$F_k(c_x^2)=L_k^{(1/2)}(c_x^2)$ and so Eq.\ (\ref{5.15}) becomes
\beq
\varphi(c_x^2)=\sum_{k=1}^\infty b_k'L_k^{(1/2)}(c_x^2).
\label{B3}
\eeq
{}From here one can obtain an alternative expression for the
coefficients $b_k'$ as
\beq
b_k'=\frac{2\Gamma(3/2)k!}{\Gamma (k+3/2)}\pi^{-1/2}
\int_{-\infty}^\infty d{c}_x\, c_x^2
e^{-c_x^2}L_k^{(1/2)}(c_x^2)\varphi(c_x^2).
\label{5.5bis}
\eeq
This is equivalent to Eq.\ (\ref{5.7bis}).


\begin{thebibliography}{99}

\bibitem{C90}
C. S. Campbell, Annu. Rev. Fluid Mech. {22} (1990) 57; I.
Goldhirsch, {ibid.} {35}  (2003) 267.

\bibitem{GvN00}
I. Goldhirsch and T. P. C. van Noije, Phys. Rev. E {61} (2000) 3241.

\bibitem{DG01}
J. W. Dufty and V. Garz\'o, J. Stat. Phys. {105} (2001) 723.

\bibitem{DBL02}
J. W. Dufty, J. J. Brey, and J. Lutsko, Phys. Rev. E {65} (2002)
051303.

\bibitem{DB02}
J. W. Dufty and J. J. Brey, J. Stat. Phys. {109} (2002) 433.

\bibitem{BDRM03}
J. J. Brey, J. W. Dufty, and M. J. Ruiz-Montero, in {Granular Gas
Dynamics}, edited by T. P\"oschel and N. Brilliantov, Springer,
Berlin, 2003, pp.\ 227--249.



\bibitem{DBB05}
J. Dufty, A. Baskaran, and J. J. Brey, J. Stat. Mech. (2006) L08002.

\bibitem{GS95}
A. Goldshtein and M. Shapiro, J. Fluid Mech. {282} (1995) 75.

\bibitem{BDS97}
J. J. Brey, J. W. Dufty, and A. Santos,  J. Stat. Phys. {87} (1997)
1051.

\bibitem{BRM04}
J. J. Brey and M. J. Ruiz-Montero, Phys. Rev. E {70} (2004) 051301.

\bibitem{BRMMG05}
J. J. Brey, M. J. Ruiz-Montero, P. Maynar, and M. I. Garc\'{\i}a de
Soria, J. Phys.: Condens. Matter  {17} (2005) S2489.

\bibitem{CC70}
S. Chapman and T. G. Cowling,{
 The Mathematical Theory of Nonuniform Gases}, Cambridge
University Press, Cambridge, 1970.

\bibitem{BDKS98}
 J. J. Brey, J. W. Dufty, C. S. Kim, and A. Santos,  Phys. Rev. E {58} (1998) 4638.

\bibitem{BC01}
J. J. Brey and D. Cubero, in {Granular Gases}, edited by T.
P\"oschel and S. Luding, Springer-Verlag, Berlin, 2001, pp.\ 59--78.

\bibitem{GM02}
V. Garz\'o and J. M. Montanero, Physica A {313} (2002) 336.

\bibitem{GD02}
V. Garz\'o and J. W. Dufty, Phys. Fluids {14} (2002) 1476; V.
Garz\'o and J. M. Montanero, Phys. Rev. E {69} (2004) 021301.

\bibitem{GD99}
V. Garz\'o and J. W. Dufty, Phys. Rev. E {59} (1999) 5895.

\bibitem{L05}
J. F. Lutsko, Phys. Rev. E {72} (2005) 021306.

\bibitem{BRMC99}
J. J. Brey, M. J. Ruiz-Montero,  and D. Cubero,  Europhys. Lett.
{48} (1999) 359.


\bibitem{GM03}
J. M. Montanero and V. Garz\'o, Phys. Rev. E {67} (2003) 021308; V.\
Garz\'o and J. M. Montanero, {ibid.} {68} (2003) 041302.

\bibitem{MSG05}
J. M. Montanero, A. Santos, and V. Garz\'o,  in {Rarefied Gas
Dynamics: 24th International Symposium on Rarefied Gas Dynamics},
edited by M. Capitelli, AIP Conference Proceedings, vol. 762,
Melville, NY, 2005, pp.\ 797--802; preprint cond-mat/0411219.

\bibitem{E82}
D. J. Evans, Phys. Lett. A {91}, 457 (1982).

\bibitem{GD83}
 M. J. Gillan and M. Dixon, J. Phys. C: Solid State Phys.
{16} (1983) 869.

\bibitem{E86}
D. J. Evans, Phys. Rev. A {34} (1986) 1449.

\bibitem{MS97}
J. M. Montanero and A. Santos,  in {Rarefied Gas Dynamics 20},
edited by C. Shen, Peking University Press, Beijing, 1997) pp.\
137--142.

\bibitem{L89}
W. Loose, Phys. Rev. A {40} (1989) 2625.

\bibitem{GS91}
V. Garz\'o and A. Santos, Chem. Phys. Lett. {177} (1991) 79.

\bibitem{GS06}
V. Garz\'o, A. Santos, and J. M. Montanero, Modified Sonine
approximation for the Navier--Stokes transport coefficients of a
granular gas, preprint cond-mat/0604079.

\bibitem{vNE98}
T. P. C. van Noije and M. H. Ernst, Gran. Matt. {1} (1998) 57.

\bibitem{EP97} S. E. Esipov and T. P\"oschel, J. Stat. Phys.
{86} (1997) 1385.


\bibitem{AS72}
{Handbook of Mathematical Functions}, edited by M. Abramowitz and I.
A. Stegun, Dover, New York, 1972.

\bibitem{MS00}
J. M. Montanero and A. Santos, Gran. Matt. {2} (2000) 53.

\bibitem{CDPT03}
F. Coppex, M. Droz, J. Piasecki, and E. Trizac, Physica A {329}
(2003) 114.

\bibitem{BD05}
J. J. Brey and J. W. Dufty, Phys. Rev. E 72  (2005) 011303.

\bibitem{DSMC}
G. Bird, {Molecular Gas Dynamics and the Direct Simulation of Gas
Flows}, Clarendon, Oxford, 1994) F. J. Alexander and A. L. Garcia,
Comp. Phys. {11} (1997) 588.


\bibitem{W92}
W. Wagner, J. Stat. Phys. {66} (1992) 1011.

\bibitem{Gallis}
M. A. Gallis, J. R. Torczynski, and D. J. Rader, Phys. Rev. E {69}
(2004) 042201.

\bibitem{BCRM99}
J. J. Brey, D. Cubero, and M. J. Ruiz-Montero, Phys. Rev. E {59}
(1999) 1256.

\bibitem{BP06}
N. V. Brilliantov and P\"oschel, Europhys. Lett. {74} (2006) 424;
Erratum: {75} (2006) 188.






\end{thebibliography}
\end{document}